\documentclass[twocolumn]{aastex63}

\usepackage{ulem}
\usepackage{amsmath}

\usepackage{multirow}
\usepackage{booktabs,tabularx,dcolumn}

\usepackage{lipsum} 
\usepackage{enumitem}
\setlist{nosep} 

\shorttitle{NGC\,3370 photometry}
\shortauthors{Jang 2022}
\graphicspath{{./}{figures/}}

\begin{document}

\title{Tests of photometry: the case of the NGC\,3370 ACS field}

\author{In Sung Jang}
\affil{Department of Astronomy \& Astrophysics, University of Chicago, 5640 South Ellis Avenue, Chicago, IL 60637}\email{isjang@astro.uchicago.edu}

\begin{abstract}
A critical analysis and comparison of different methods for obtaining point spread function (PSF) photometry are carried out.
Deep ACS observations of NGC\,3370 were reduced using four distinct approaches.
These reductions explore a number of methodological differences: software packages (DAOPHOT and DOLPHOT), input images (individual and stacked frames), PSF models (synthetic and empirical), and aperture correction methods (automatic and manual).
A comparison of the photometry leads to the following results:
1) Photometric incompleteness between individual reductions shows only a minimal difference ($<$10\%). 
2) Statistical errors are 20\% to 30\% smaller for DAOPHOT runs on stacked frames than DOLPHOT runs on individual frames.
3) Statistical errors assigned directly by the photometry codes are 25\% to 50\% smaller than the errors measured from artificial star tests.
4) Systematic errors are magnitude dependent and become larger at the faint end, at the level of $\sigma_s\sim0.1$~mag.
5) The automatic aperture correction routines in DOLPHOT result in a significant systematic error ($\sigma_s \sim 0.05$~mag).
6) Individual reductions agree well at the 0.02 mag level when the systematic errors are properly corrected through artificial star tests. 
The reasonable agreement between the reductions leads to important implications that i) the reduction dependent errors can be reduced to a 1\% level in the luminosity distance scale, and ii) the stacked frame photometry can be a good means to study non-variable stars in external galaxies.

\end{abstract}


\section{Introduction}
Measuring accurate fluxes of resolved stellar objects is one of the most fundamental steps in observational astronomy.
This is due to the nature of astronomy---the information content of stellar objects comes primarily from their electromagnetic radiation. 
Measuring this radiation is therefore the first task in deriving stellar physical properties (e.g. temperature) that can be used to constrain stellar evolution models.
When stars are observed in nearby resolved galaxies this flux can then be combined with stellar distance indicators (i.e., standard candles) to determine their accurate luminosity distances, which can then form the basis for determining cosmological parameters, such as the Hubble constant ($H_0$).

There have been increasing efforts to determine the local value of the Hubble constant using stellar distance indicators.
And, remarkable progress has been made in the past few years:

\setlist[2]{noitemsep} 
\setenumerate{noitemsep} 
\begin{enumerate}[noitemsep] 
\item Geometric distances to the local calibrator galaxies have increased in precision (LMC -- \citet{pie19}, SMC -- \citet{gra20}, NGC\,4258 -- \citet{rei19}, Milky Way calibrators -- \citet{gai21});
\item The Tip of the Red Giant Branch (TRGB) method has been used to determine the distances to Type Ia supernova (SN Ia) host galaxies \citep[e.g.,][]{fre19, fre20}, providing a measurement of $H_0$ that is independent of the Cepheid distance scale.
\item A larger number of local SN~Ia calibrators and host galaxies have been observed (19 SNe Ia -- \citet{fre21} and 42 SNe Ia -- \citet{rie21}).
\end{enumerate}
The quoted accuracy of the Hubble constant measured from stellar distance indicators is now nearing $\sim$1\% precision, corresponding to only $\sim$0.02 mag in the luminosity scale.

The improved accuracy of the local distance scale is encouraging.
However, it should be noted that 
these measurements rest on an uncertain interface, the photometry.
For example, the mean photometric error 
for a single TRGB star at $\sim$20~Mpc is $\sigma_{F814W} \sim0.15$~mag
in $HST$ imaging \citep[e.g.,][]{jan18}.
For Cepheid variables, the dispersion in the NIR period-luminosity relation is on average $\sim$0.3~mag for galaxies in the same distance range \citep[e.g.,][]{rie16}.
These individual errors are much larger than the final error of the Hubble constant.
While statistical errors can be reduced 
by analyzing a larger number of stars, 
there may be systematic issues that can not be reduced by increasing the sample size and are more difficult to identify and constrain.

Stellar photometry is an art that requires stringent control of errors \citep[e.g., ][]{ste87}.
The theory of stellar photometry is straightforward: one must determine the local sky background and then measure the flux excess using a fixed aperture (aperture photometry) or a pre-calculated Point-Spread-Function model (PSF fitting photometry).
However, from the very first step, the practical implementation is complex, and many choices must be made.
For example, there is no single sky annulus that is optimal for all degrees of stellar crowding.
Even using a given sky annulus, 
a number of statistical schemes can be considered to find a representative value of the sky background: $mean$, $median$, and $mode$\footnote{$mode$ is not necessarily unique to the discrete distribution so its approximation can also be considered: $mode \approx 3\times median - 2\times mean$ (e.g., DAOPHOT).}, and one may further consider different values for the sigma clipping of outlier pixels. 
In a crowded field, unresolved sources contribute to the sky background, making a gradient within the sky annuli.
These issues are all possible sources of error.

Artificial star tests have been the preferred method to assess the robustness of crowded-field photometry \citep[e.g., ][]{ste88}.
These tests are performed by injecting a large number of artificial stars into the images and then recovering them in the same manner as real stars.
This process is computationally expensive, but it does deliver robust estimates of photometric performance (e.g., completeness, random errors, and systematic errors). 
However, it  does not fully encapsulate the all potential sources of error.
One example is the error associated with the use of incomplete PSF models.
Model PSFs do not exactly match the intrinsic image PSFs, so there is an uncertainty in the measured flux of real sources due to the difference in PSFs. It is not possible to measure this bias from artificial star tests alone, since the artificial stars are injected using the model PSF, which differs from the real sources.

A complementary and alternative approach is to undertake and compare independent reductions, 
 from different input parameter sets, 
or ideally from different software packages \citep{fre01}. 
Indeed, such tests have been carried out in the literature\footnote{For $HST$ images, we identified the following studies: \citet{kel96, sil96, phe98, kel99, sil99, dol00, rej05, mon10, hat17, jan17a, jan18, fil20, jan21}. 
}.
These studies often found small, but non-negligible magnitude-dependent offsets, such that the faint stars show larger discrepancy ($\Delta mag \lesssim 0.05$).
The faint side of photometry is a sensitive test area for systematic effects. 
The observed discrepancy could be due to the different photometric processing or the systematic bias associated with the stellar crowding and sky subtraction. 
The latter can be quantified through artificial star tests.
Therefore, it would be ideal to combine both approaches---artificial star tests and independent reductions---to address  systematic issues in photometry whenever practical.

\begin{figure}
\centering
\includegraphics[scale=0.8]{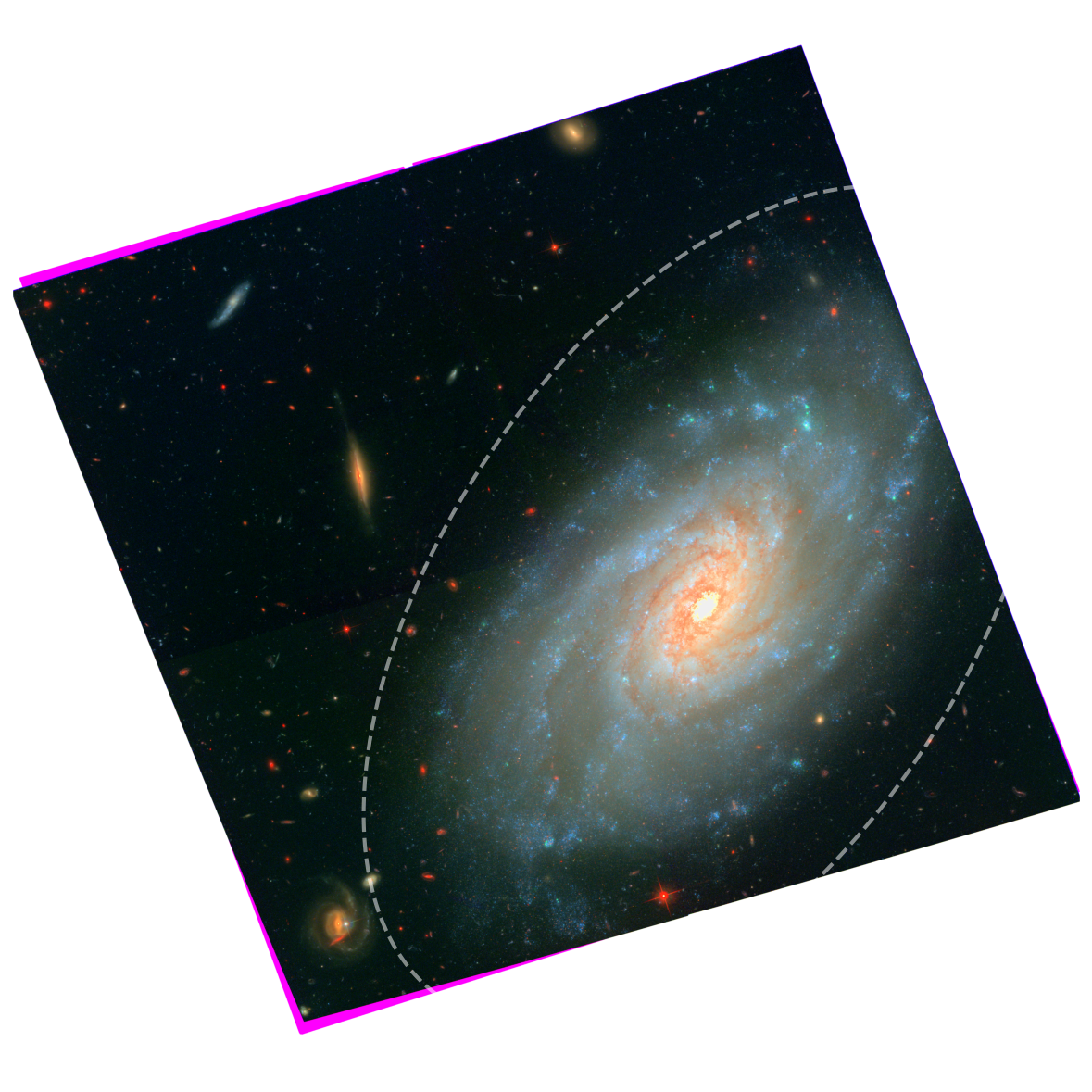}
\caption{Color image of the NGC\,3370 ACS field.
Red, green and blue channels are taken from F814W, F555W, and F435W, respectively.
North is up and east is to the left.
The dashed ellipse indicates the boundary between the inner and outer regions used in this study ($SMA = 2\arcmin$).
}
\label{fig1}
\end{figure}

In this paper, we now present detailed photometry tests applied to deep observations of NGC\,3370 taken with ACS/WFC on board the Hubble Space Telescope ($HST$).
NGC\,3370 is a moderately inclined late-type disk galaxy (Figure \ref{fig1}) and has been host to a Type Ia Supernova, SN\,1994ae.
The ACS observations for this galaxy provide a unique opportunity for this study since two compelling stellar distance indicators have been applied to the $same$ $dataset$ and resulted in accurate distances: 
Cepheids \citep[][$(m-M)_0 = 32.29\pm0.06$]{rie05} and 
the Tip of the Red Giant Branch (TRGB) \citep[][$(m-M)_0 = 32.25\pm0.04$]{jan17b}.
We reduced this ACS dataset in a number of independent ways, thereby allowing us to compare photometric performance and to test possible reduction-dependent uncertainties.

This paper is organized as follows.
Section~2 describes the ACS imaging data and data reduction procedures.
We provide details of how we obtained PSF photometry from the individual reductions, and how we carried out aperture corrections and artificial star tests.
Section~3 presents a detailed comparison of photometric performance (e.g., completeness, precision, and accuracy) measured from artificial stars.
Furthermore, we examine magnitudes of real stars to see if there are any reduction-dependent systematics.
The primary results are summarized in Section~4.

\section{Data and Data processing}
%%%%%%%%%%%%%%%%%%%%%%%
% Figure 1
%%%%%%%%%%%%%%%%%%%%%%%

\subsection{Archival Image Data}
The images for NGC\,3370 used in this study were downloaded from the MAST archive.
Observations were taken with the ACS/WFC instrument on $HST$ in three filters: $F435W$ (9,600s), $F555W$ (61,240s), and $F814W$ (24,000s).
The $F435W$ data\footnote{Obtained as part of the Hubble Heritage Program (PID: 9696, PI: K. Noll).} were only used to make the color composite image shown in Figure \ref{fig1}, and not used in the subsequent analysis.
The $F555W$ and $F814W$ data were originally aimed at detecting Cepheid variables (PID: 9351 and 10802, PI: A. Riess), and here we use them to carry out photometry tests.
All of the observations, except for some short $F555W$ exposures (3,640s), were taken in 2003. 
This means that the dataset is one of early applications of ACS (installed in 2002), when the instrument had better sensitivity and lower CTE losses.

Individual frames were processed before extracting the photometry, following the methods described in \citet{jan21}. 
To briefly summarize, we carried out initial PSF photometry on the individual \texttt{\_flc} images, selecting bright stellar objects. These objects were then used as the alignment sources for Tweakreg in DrizzlePac 2.0 \citep{gon12, avi15}.
This process found that some images were not properly aligned 
with a measurable offset of $\sim$1~pixel with respect to the reference frame.
The aligned \texttt{\_flc} images were then passed into Astrodrizzle to generate stacked \texttt{\_drc} images. 
We prepared one drizzled image from the individual $F814W$ images to serve as the reference frame for the DOLPHOT reductions (Phot\,A and B, see next section for details) using the default pixel scale of $0\farcs05$. 
Additionally, we prepared a drizzled image in each $F555W$ and $F814W$ to perform drizzled frame photometry in DAOPHOT (Phot\,C and D). These images were generated using a slightly finer pixel scale of 0\farcs03, which mitigates the undersampling of the ACS point spread function (PSF).
All drizzled images were generated using a \texttt{final\_pixfrac} of 1 and \texttt{final\_kernel} of `\texttt{square}' (default).

\subsection{Point Spread Function (PSF) Photometry}

\begin{deluxetable*}{cccccc} %%%%%%%%%%%%%%%%%%%%%%%%%%%%%
\tabletypesize{\small}
\setlength{\tabcolsep}{0.05in}
\tablecaption{Summary of photometric reduction methods \label{tab1}}
\tablewidth{0pt}
\tablehead{\colhead{Processing ID} & \colhead{Photometry tool} & \colhead{Input image} & \colhead{PSF model} & \colhead{Aperture correction} &\colhead{Sky estimation}}
\startdata
Phot\,A & DOLPHOT   & Individual frame (\texttt{\_flc}) & TinyTim & Auto (\texttt{ApCor=1}) & \texttt{Fitsky=2}\\ 
Phot\,B & DOLPHOT   & Individual frame (\texttt{\_flc}) & TinyTim & Manual & \texttt{Fitsky=2}\\ 
Phot\,C & DAOPHOT   & Drizzled frame (\texttt{\_drc}) & Empirical & Manual & ...\\ 
Phot\,D & DAOPHOT   & Drizzled frame (\texttt{\_drc}) & TinyTim & Manual & ...\\ 
\enddata
\end{deluxetable*} %%%%%%%%%%%%%%%%%%%%%%%%%%%%%

\begin{deluxetable}{lll} %%%%%%%%%%%%%%%%%%%%%%%%%%%%%
\tabletypesize{\footnotesize}
\setlength{\tabcolsep}{0.05in}
\tablecaption{DOLPHOT Processing Parameters \label{tab2}}
\tablewidth{0pt}
\tablehead{\colhead{Processing ID} & \colhead{Parameter} & \colhead{Value}  }
\startdata
All & \texttt{PSFPhot}     & 1  \\ 
All & \texttt{PSFPhotIt}   & 2  \\ 
All & \texttt{Force1}      & 0  \\ 
All & \texttt{SkipSky}     & 2    \\   
All & \texttt{SkySig}      & 2.25 \\   
All & \texttt{SecondPass}  & 5    \\   
All & \texttt{SearchMode}  & 1     \\  
All & \texttt{SigFind}     & 2.5   \\  
All & \texttt{SigFindMult} & 0.85 \\   
All & \texttt{SigFinal}    & 2.5   \\  
All & \texttt{MaxIT}       & 25   \\   
All & \texttt{NoiseMult}   & 0.10 \\   
All & \texttt{FSat}        & 0.999 \\  
All & \texttt{RCentroid}   & 2 \\      
All & \texttt{PosStep}     & 0.25 \\  
All & \texttt{dPosMax}     & 3.0   \\  
All & \texttt{RCombine}    & 1.5 \\  
All & \texttt{SigPSF}      & 3.0 \\    
All & \texttt{PSFres}      & 1 \\      
All & \texttt{useWCS}      & 1 \\  
All & \texttt{Align}       & 2 \\  
All & \texttt{Rotate}      & 1 \\  
All & \texttt{FlagMask}    & 4 \\  
All & \texttt{InterpPSFlib}& 1 \\ 
All & \texttt{ACSpsfType}  & 0 \\ 
All & \texttt{ACSuseCTE}   & 0 \\
All & \texttt{FitSky}      & 2    \\ 
All & \texttt{img\_RAper}  & 3    \\
All & \texttt{img\_RChi}   & 2  \\
All & \texttt{img\_RPSF}   & 15 \\
All & \texttt{img\_RSky}   & 15 35 \\
All & \texttt{img\_RSky2}  & 4 10 \\
All & \texttt{img\_apsky}  & 15 25 \\ 
\hline
Phot\,A & \texttt{ApCor}       & 1    \\      
\hline
Phot\,B & \texttt{ApCor}       & 0    \\     
\enddata
\end{deluxetable} %%%%%%%%%%%%%%%%%%%%%%%%%%%%%

There are two possible approaches to extracting photometry from multiple images of the same field:
1) Use individual images and photometer them simultaneously.
The mean magnitude is then determined from the individual flux measurements.
2) Make a co-added (stacked) frame in each band and perform the photometric measurements on the stack.

The first approach allows one to generate time-series photometry 
of variable stars.
Because input images are only passed through the basic processing steps (e.g., bias correction, flat fielding), the intrinsic image PSF is preserved.
This helps in achieving high photometric accuracy when the image PSF is well characterized by synthetic models, as is the case with $HST$ data.
DOLPHOT is a modified version of HSTphot \citep{dol00} and has a number of useful routines to simultaneously photometer individual $HST$ images using  synthetic PSFs (e.g., TinyTim PSFs).
It has been widely used in recent studies \citep[e.g., ][]{tul09, dal09, rad11, ana21} and  it was also tested in this study.

The second approach has been considered to be sub-optimal for accurate photometry 
because the pixel counts are resampled during the stacking process.
However, improved image processing techniques in recent years  (e.g., DrizzlePac) have made stacked frame photometry more robust.
This approach requires significantly reduced computing resources than photometering the individual images simultaneously.
In general, empirical PSFs, which can be constructed from bright stars in the stacked frame, are used to carry out PSF photometry (but see Phot\,D, as described below and in Table \ref{tab1}.).
Stacked frame photometry has been used to reduce $HST$ data in many studies\footnote{For ACS/WFC, the following studies can be found: 
\citet{pir05, rej05, bro06, cal06, alo07, wil07, sav08, bro09, bir10, kal12, ann13, bro14, geh15, lee16b, jan17b, tik19, fil20}} and it is more common in ground-based surveys (e.g., Dark Energy Survey \citep{bur18}, Hyper Suprime-Cam Subaru Strategic Program \citep{aih18}).

%%%%%%%%%%%%%%%%%%%%%%%
% Figure 2
%%%%%%%%%%%%%%%%%%%%%%%
\begin{figure*}
\centering
\includegraphics[scale=1]{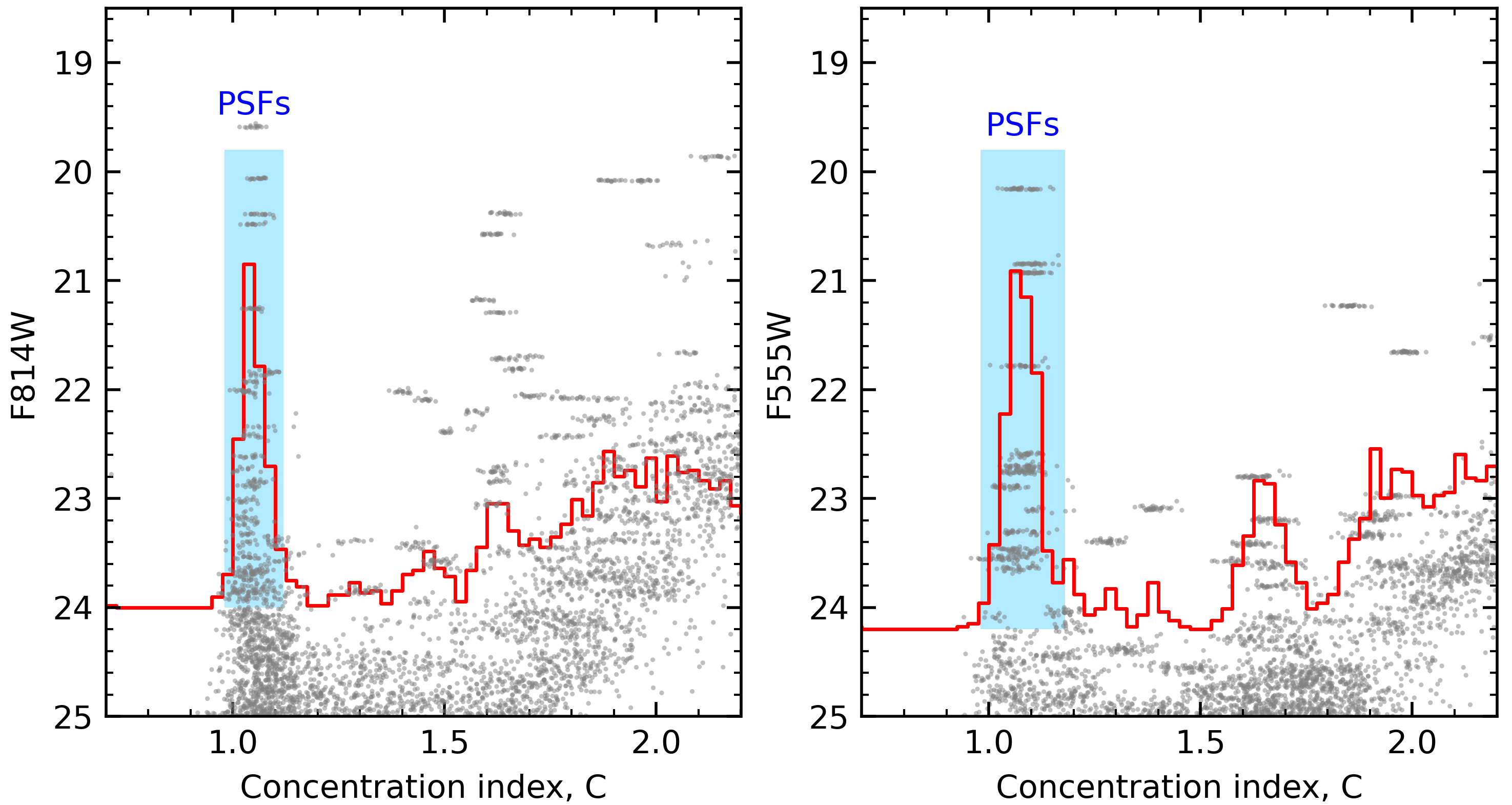}
\caption{Selection of point sources for the aperture correction in $F814W$ (left) and $F555W$ (right).
We used sources in the outer region with  $SMA > 1\farcm3$, to avoid high stellar crowding. 
The concentration index ($C$) is defined as the magnitude difference (in \texttt{\_flc}) between the small and large aperture radii: $C = m(r=0\farcs04) - m(r=0\farcs125)$. 
The red line is a histogram for sources brighter than 24~mag in each filter.
A strong plume of point sources is clearly seen at $C\sim1.1$.
Sources in the cyan area were used for the aperture correction after a careful visual inspection.
}
\label{fig2}
\end{figure*}

In this study, we tested both approaches with a number of parameter changes. 
The details are listed below and also summarized in Table \ref{tab1}.\\

\begin{itemize}
    \item Phot\,A: 
    uses DOLPHOT with individual frame \texttt{\_flc} images. 
    The input parameters are listed in Table \ref{tab2}.
    They are consistent with those in the DOLPHOT/ACS User’s Guide, except for \texttt{SigFinal=2.5}.
    This is slightly lower than the default value (\texttt{3.5}) and makes sure that all of the detected stars (\texttt{SigFind=2.5}) are listed in the final catalog.
    The PSF to $0\farcm5$ aperture correction is measured using the DOLPHOT implemented routines (\texttt{ApCor=1}). 
    The local sky background is measured with \texttt{Fitsky=2}.
    This is 
    relevant to most of the DOLPHOT processing in the literature. \\

    \item Phot\,B: same as the Phot\,A processing, but with manual aperture corrections (see Section~2.3 for details).\\

    \item Phot\,C: uses drizzled stacked frames (\texttt{\_drc}) to measure stellar flux. 
    A single pass of DAOPHOT PSF photometry (\texttt{FIND - PHOT - ALLSTAR - ALLFRAME}) is carried out \citep{ste87, ste94}. 
    We used a readout noise of 4.98e$^-$ and a gain of 1.0e$^-$/ADU, as listed in the header of the drizzled images.
    PSF images were constructed empirically using $\sim$20 isolated, bright stars in the drizzled frames. 
    Source coordinates are determined from the montaged frame with a detection threshold of $\sim$3$\sigma$, where the sky fluctuation is measured in the outskirts of the galaxy ($SMA\sim3\arcmin$).\\
    
    \item Phot\,D: same as the Phot\,C processing, except that the TinyTim PSFs were used. 
    We used the "\texttt{addstars}" task in DOLPHOT to inject TinyTim PSFs on individual \texttt{\_flc} images and then drizzled them in the same way as the original images. 
    We note that the injected PSFs are corrected for the difference between the image PSFs and the models, measured from the prior DOLPHOT run (i.e., central pixel adjustment).
    This increases the photometric accuracy.\\
\end{itemize}

There are three local sky fitting options generally used in DOLPHOT: \texttt{Fitsky=1}, \texttt{2}, and \texttt{3}.
The \texttt{Fitsky=1} option measures the sky background prior to each photometry measurement with an annulus given by  \texttt{img\_RSky} (\texttt{=15} and \texttt{35} pixels).
Sky measurements in DOLPHOT are sigma-clipped $means$, with \texttt{SkySig} (\texttt{=2.25}) being the sigma used for the clipping.
The \texttt{Fitsky=2} option fits the sky first, and the star second (i.e., two single-parameter fits). 
The sky annulus is given by \texttt{img\_RSky2} (\texttt{=4} and \texttt{10} pixels), typically smaller than \texttt{img\_RSky} used for \texttt{Fitsky=1} and \texttt{3}.
The \texttt{Fitsky=3} option makes a single fit for the sky and star simultaneously (i.e., one two-parameter fit). 
In theory, \texttt{Fitsky=3} with \texttt{img\_RAper=10} should be very similar to \texttt{Fitsky=2} with \texttt{img\_RAper=3} and \texttt{img\_RSky2=4 10}, but \texttt{Fitsky=2} is significantly more robust in extremely crowded fields (A. Dolphin 2021, private communication). % (TBA).
In this study, we chose the \texttt{Fitsky=2} option to simultaneously process the inner crowded and outer sparse regions of NGC\,3370.

DAOPHOT measures the local sky background from the $mode$ of pixel counts after sigma clipping. 
Here the $mode$ is an approximate estimate with an empirical relationship: $mode = 3\,\times median - 2\, \times  mean$.

%%%%%%%%%%%%%%%%%%%%%%%
% Figure 3
%%%%%%%%%%%%%%%%%%%%%%%
\begin{figure*}
\centering
\includegraphics[scale=0.7]{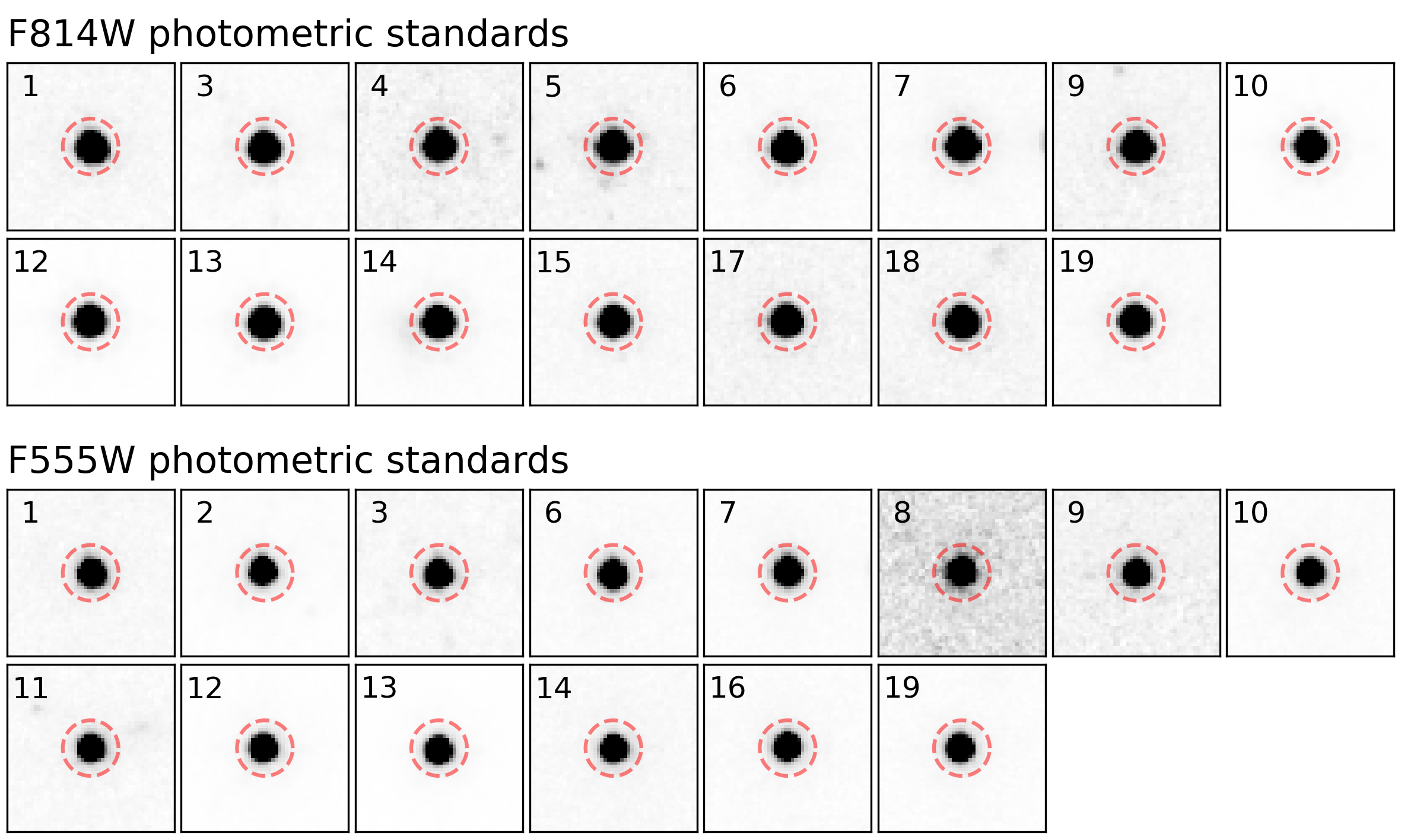}
\caption{$1\farcs5 \times 1\farcs5$ thumbnail images of the photometric standards in $F814W$ (top) and $F555W$ (bottom). Dashed circles indicate the $r=0\farcs25$ radius we used for the aperture photometry. Identification numbers are taken from Table \ref{tab3}.
}
\label{fig3}
\end{figure*}

 \begin{deluxetable*}{ccc|cccccccc|cc} %%%%%%%%%%%%%%%%%%%%%%%%%%%%%
\tabletypesize{\small}
\setlength{\tabcolsep}{0.05in}
\tablecaption{A List of the Photometric Standards \label{tab3}}
\tablewidth{0pt}
\tablehead{\colhead{ID} & \colhead{R.A.} & \colhead{Decl.} & \multicolumn{8}{c}{FLC ($r=0\farcs25$)}  & \multicolumn{2}{c}{DRC ($r=0\farcs25$)} \\
& & & \colhead{$F814W$} & \colhead{Err} &\colhead{N} &\colhead{Stdev} & \colhead{$F555W$} & \colhead{Err} &\colhead{N} &\colhead{Stdev} & \colhead{$F814W$} & \colhead{$F555W$}}
\startdata
1  & 161.740256 & 17.261232 &  22.340 &  0.009 &   8  &  0.040 &  23.103 &  0.014 & 21  &  0.013  & 22.334  & 23.096  \\
2  & 161.751915 & 17.281692 &     ... &   ...  &  ... &   ...  &  21.748 &  0.006 & 36  &  0.011  &   ...   & 21.757  \\
3  & 161.753383 & 17.270955 &  22.013 &  0.008 &  16  &  0.012 &  23.555 &  0.017 & 32  &  0.015  & 22.013  & 23.543  \\
4  & 161.755169 & 17.263363 &  23.174 &  0.016 &  13  &  0.027 &   ...   &   ...  & ... &   ...   & 23.153  &  ...    \\
5  & 161.767836 & 17.294145 &  22.887 &  0.013 &  13  &  0.013 &   ...   &   ...  & ... &   ...   & 22.889  &  ...    \\
6  & 161.775137 & 17.291700 &  21.929 &  0.007 &  12  &  0.016 &  22.703 &  0.010 & 40  &  0.017  & 21.932  & 22.708  \\
7  & 161.780718 & 17.303599 &  21.854 &  0.007 &  14  &  0.013 &  22.592 &  0.009 & 31  &  0.016  & 21.859  & 22.570  \\
8  & 161.782637 & 17.286926 &    ...  &    ... &  ... &   ...  &  24.368 &  0.055 &  7  &  0.048  &  ...    & 24.404  \\
9  & 161.783071 & 17.286043 &  22.745 &  0.012 &   6  &  0.051 &  24.088 &  0.030 & 15  &  0.032  & 22.741  & 24.092  \\
10 & 161.789391 & 17.250907 &  20.392 &  0.003 &  16  &  0.004 &  22.736 &  0.010 & 31  &  0.015  & 20.387  & 22.729  \\
11 & 161.791513 & 17.259049 &    ...  &    ... &  ... &   ...  &  23.460 &  0.016 & 35  &  0.017  &  ...    & 23.475  \\
12 & 161.792929 & 17.263690 &  20.481 &  0.003 &  13  &  0.006 &  21.784 &  0.005 & 36  &  0.018  & 20.475  & 21.782  \\
13 & 161.794224 & 17.267171 &  20.063 &  0.003 &  16  &  0.005 &  20.927 &  0.004 & 47  &  0.007  & 20.060  & 20.926  \\
14 & 161.798045 & 17.265244 &  21.258 &  0.005 &  15  &  0.005 &  ...    &   ...  & ... &   ...   & 21.252  &  ...    \\
15 & 161.801675 & 17.260026 &  22.420 &  0.009 &  14  &  0.020 &  23.307 &  0.014 & 39  &  0.023  & 22.416  & 23.306  \\
16 & 161.802506 & 17.289145 &    ...  &    ... &  ... &   ...  &  22.894 &  0.011 & 34  &  0.016  &  ...    & 22.904  \\
17 & 161.802663 & 17.283882 &  23.017 &  0.014 &  15  &  0.016 &  ...    &   ...  & ... &   ...   & 23.013  &  ...    \\
18 & 161.803457 & 17.263566 &  22.845 &  0.012 &  15  &  0.013 &  ...    &   ...  & ... &   ...   & 22.841  &  ...    \\
19 & 161.806155 & 17.291105 &  22.029 &  0.007 &   8  &  0.011 &  22.761 &  0.010 & 36  &  0.016  & 22.014  & 22.758  \\
\enddata
\end{deluxetable*} %%%%%%%%%%%%%%%%%%%%%%%%%%%%%

\subsection{Aperture correction}

The finite sizes of PSF models and sky annuli used for photometry require an additional flux correction which is called the ``aperture correction".
For the photometry of $HST$ data, the correction procedure is divided into two steps: 
1) a correction from the PSF fit magnitude to an aperture magnitude at finite radius, and 
2) a correction from the finite aperture magnitude to infinity.
The second step can be made with the encircled energy values provided by STScI, so this section focuses on the first step, an empirical determination of the aperture correction out to a finite radius.

The general procedure of the aperture correction is to select bright and isolated stars (hereafter, photometric standards) and then compare their aperture magnitudes with the PSF fit magnitudes. 
The mean magnitude difference in each band (or in each exposure) is identified as the aperture correction.
DOLPHOT provides an automated routine to do the necessary correction (\texttt{ApCor=1}), which is used in Phot\,A.
For the other reductions (Phot\,B -- D), we applied the aperture correction determined manually, following the method described in \citet{jan21}.

Figure \ref{fig2} displays the selection of bright point sources we used for the aperture correction.
The concentration index, $C$, is the magnitude difference between the small and large aperture radii.
This simple parameter is very efficient at distinguishing between point sources and marginally resolved sources \citep[e.g.,][]{whi99}, as shown in the study of globular clusters in the Coma cluster \citep{lee16a}.
We derived the concentration index from the photometry of individual \texttt{\_flc} images 
with aperture radii of $r=0\farcs04$ and $0\farcs125$.
The histogram of the detected sources (red line) shows a prominent peak at $C\sim1.05$, which we identified as point sources.
We selected 15 and 14 bright point sources in the $F814W$ and $F555W$ bands respectively, after a careful visual inspection of images and the light growth curves.

Figure \ref{fig3} shows the selected bright stars in the drizzled frame. 
Their  $r=0\farcs25$ aperture magnitudes are listed in Table \ref{tab3}.
Photometry was carried out on both individual (\texttt{\_flc}) and drizzled (\texttt{\_drc}) frames.
We found that there is no significant difference between the two image sets.
The mean (median) difference in $F814W$ is 0.0045 (0.0040) mag with a standard deviation of 0.0063 mag.
The same calculation in $F555W$ gives a mean (median) difference of --0.0017 (0.0015) mag with a standard deviation of 0.0132 mag. 
This good agreement in aperture photometry indicates that the stellar flux is well preserved after resampling images with DrizzlePac. 

With the $r=0\farcs25$ aperture magnitudes in hand, we calibrated the PSF fit magnitudes.
Here we separated the image types for the correction: PSF fit magnitudes from the individual frames (Phot\,A and B) or drizzled frames (Phot\,C and D) were combined with aperture magnitudes from the same image frames. 
The corrections from the finite aperture magnitudes ($r=0\farcs25$) to infinity were made with those from \citet{boh16} : --0.1726~mag for $F814W$ and --0.1537~mag for $F555W$.
We adopt photometric zero-points from the webtool\footnote{https://acszeropoints.stsci.edu/} provided by STScI: 25.525 for $F814W$ and 25.725 for $F555W$.

%%%%%%%%%%%%%%%%%%%%%%%
% Figure 4
%%%%%%%%%%%%%%%%%%%%%%%
\begin{figure}
\centering

\includegraphics[scale=0.7]{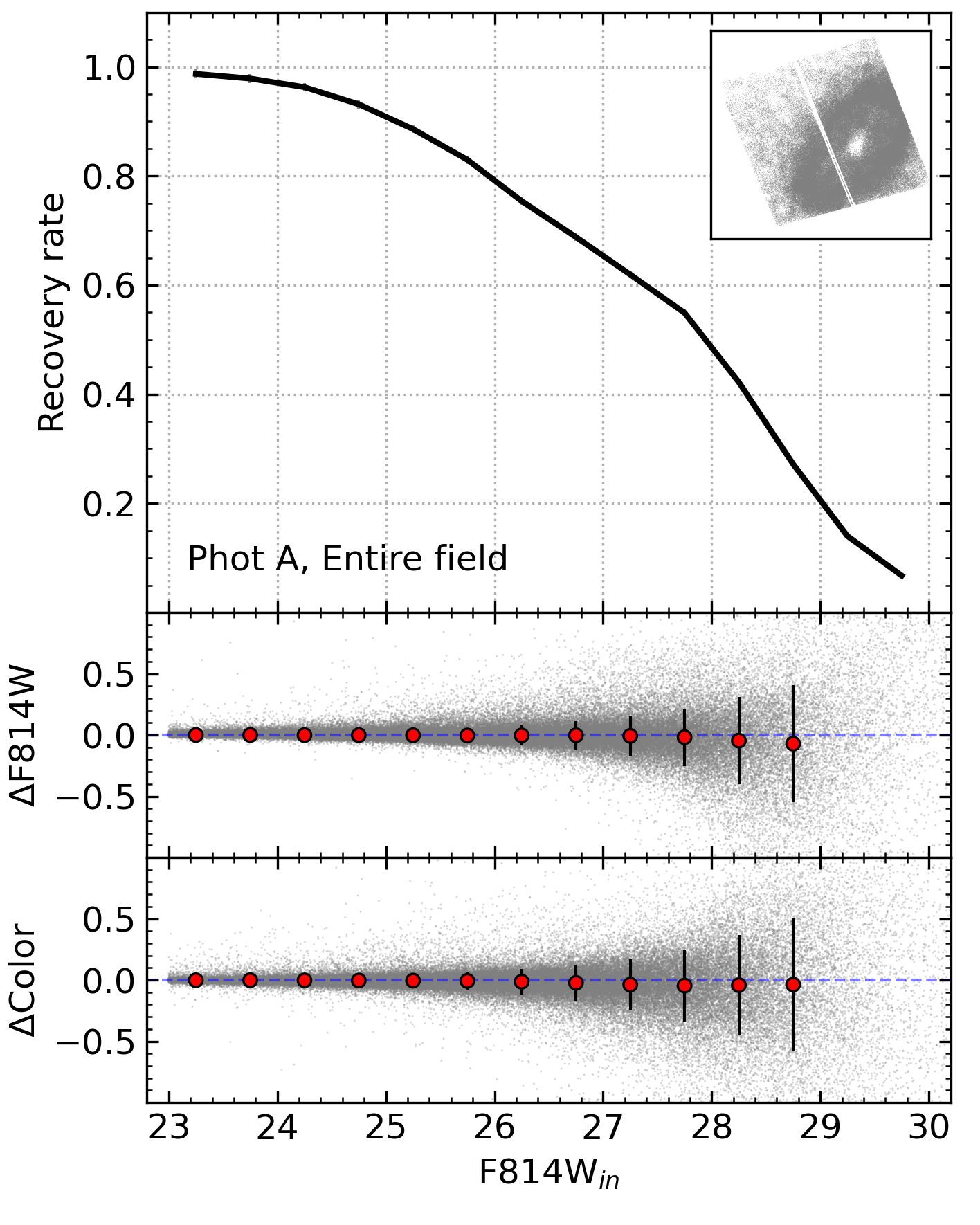}
\caption{Results of artificial star tests for Phot\,A. 
(top) Recovery rates as a function of $F814W$ magnitude.
Error bars are mostly smaller than the line width.
The inset represents the spatial distribution of artificial stars.
(middle) Differences between the injected and recovered stars ($\Delta F814W= F814W_{in} - F814W_{out}$).
Median differences and standard deviations of in each magnitude bin are marked by red dots and error bars, respectively.
(bottom) Same as the middle panel, but for $F555W - F815W$ colors. 
}
\label{fig4}
\end{figure}

\subsection{Artificial star tests}
Extensive artificial star tests have been carried out to estimate errors and completeness of our photometry.
We made a list of $5\times10^5$ artificial stars that has a uniform distribution in the color range of 
$0.5 < F555W - F814W < 2$ 
and the magnitude range of $23.0 < F814W < 30.5$~mag.  
The spatial distribution of the artificial stars was set to mimic the real stars: we selected real stars with $F814W \leq 29$~mag and used their coordinates with random shifts of
$|\Delta| \leq 5\arcsec$ along the $X$ and $Y$ directions. 
We used the same list of the artificial stars for Phot\,A -- D.

The artificial stars were injected into images and recovered in the same way as was done for the real stars. 
For the DAOPHOT-based processing (Phot\,C and D), 
we injected a small number of artificial stars ($N \sim 10,000$) at a time to preserve the degree of stellar crowding. 
Instead, we repeated the test for 50 times to sample the full list of the artificial stars ($N = 5\times10^5$).
In the case of the DOLPHOT-based processing (Phot\,A and B), such an issue is not present, 
because the code performs the test one star at a time.

Figure \ref{fig4} shows the results of the artificial star tests for Phot\,A.
We used sources that passed the point source selection criteria (details are given in the next section).
We found that the photometry is moderately deep with a 50\% recovery rate at $F814W \sim 28.1$~mag (top panel). 
The mean differences in $F814W$ magnitude and $(F555W-F814W)$ color between the injected and recovered stars are measured to be small ($<0.05$~mag) for $F814W < 29$~mag (middle and bottom panels).
In Section~3 we explore how the photometric errors and completeness vary depending on the spatial selection, color selection, and the data reduction methods (Phot\,A -- D).

%%%%%%%%%%%%%%%%%%%%%%%
% Figure 5
%%%%%%%%%%%%%%%%%%%%%%%
\begin{figure}
\centering
\includegraphics[scale=0.9]{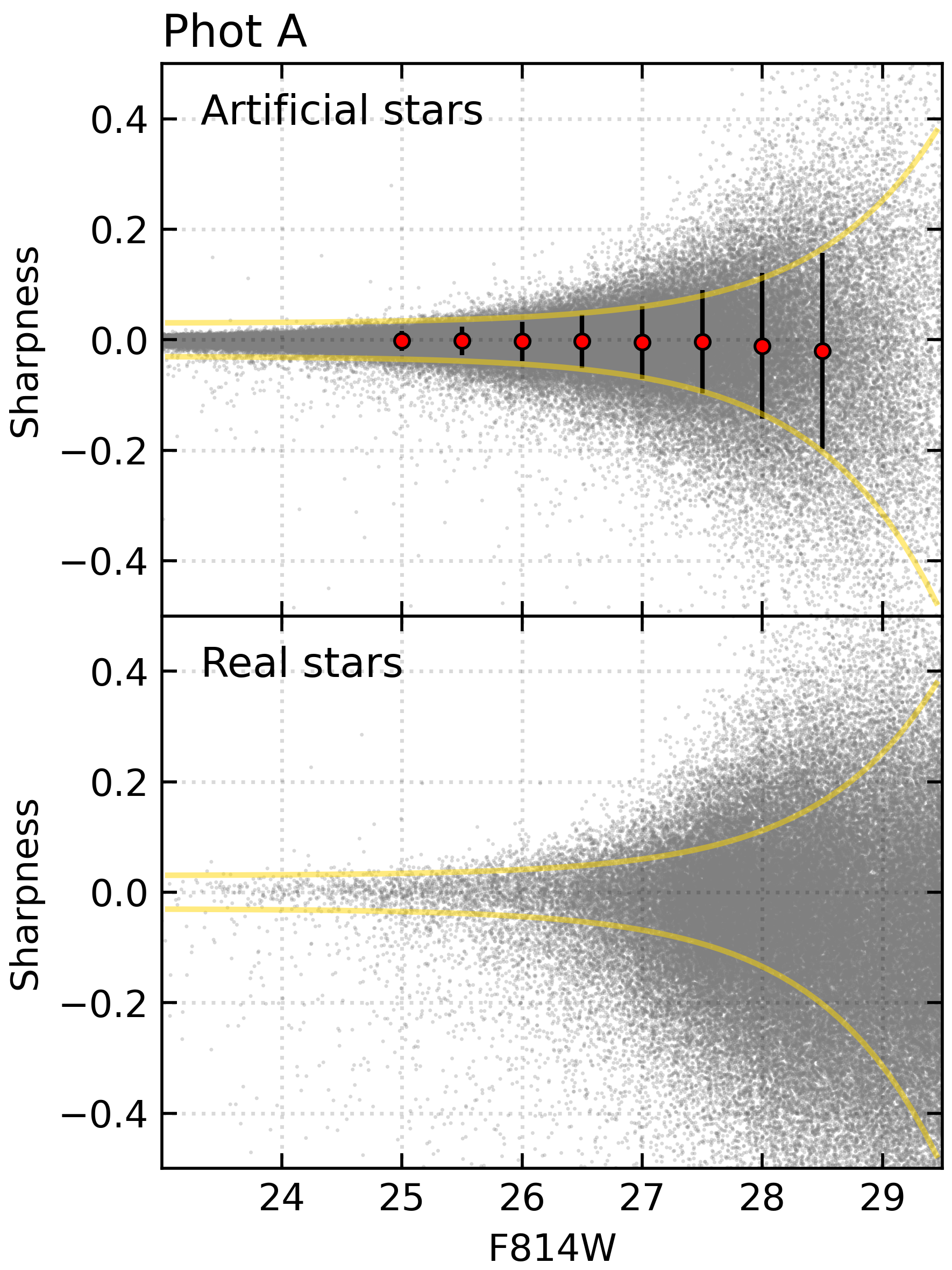}
\caption{Selection of point sources for Phot\,A. 
(Top) sharpness vs. $F814W$ magnitudes of artificial stars in the region with $SMA \geq 1\arcmin$.
Red circles with error bars represent the median and standard deviation in each magnitude bin, respectively.
The upper and lower boundaries of the error bars were fitted with a constant+exponential function (yellow lines). 
(Bottom) Same as the top panel, but for the real stars. 
Yellow lines are taken from the top panel.
Sources between the two lines were considered as point sources.
}
\label{fig5}
\end{figure}

\subsection{Point source selection}

There are several types of sources in the raw photometry catalogs of the NGC\,3370 field: 
stars, spatially-resolved star clusters, blended sources, background galaxies, and false stellar detections. 
Selecting reliable point sources ($\approx$ stars) is one of the main steps before comparing PSF photometry. 
We inspected photometric diagnostic parameters returned from DOLPHOT and DAOPHOT and 
chose the sharpness parameter as our criterion for the point source selection.

Figure \ref{fig5} displays an illustration of our point source selection criterion applied to Phot\,A.
We used the sharpness distribution of artificial stars 
as a proxy for the genuine point sources (top panel).
Red dots indicate the median sharpness values in each magnitude bin.
Their standard deviations are marked by error bars.
We determine the point source selection criteria by fitting the upper and lower ends of the error bars (yellow lines) with a constant + exponential function as follows:

\begin{equation}
{\rm Sharpness}_{F814W} = \alpha + \beta\times{\rm exp}(F814W - \gamma).
\end{equation}

\noindent Here we fit the upper and lower boundaries separately, because the sharpness distributions are not always symmetric.
For the upper boundary, we fixed $\alpha = 0.03$, and found $\beta = 0.00023$ and $\gamma = 22.08$.
The same scheme for the lower boundary gives  $\alpha = -0.03$, $\beta = -0.00032$ and $\gamma = 22.15$.
We then applied these selection criteria to the real stars (bottom panel). 
Sources between the two boundary lines were considered as point sources. 
We applied the same methodology to other reductions (Phot\,A--D). 
The derived values are listed in Table~\ref{tab4}.

In addition to the sharpness-based selection above, we applied \texttt{type} = 1 (good star) for  Phot\,A and B.
This additional cut effectively eliminates spurious detections in the DOLPHOT processing.
DAOPHOT does not provide the \texttt{type} parameter, so we applied the sharpness-based cut only (Phot\,C and D).

\begin{deluxetable}{ccccccc} %%%%%%%%%%%%%%%%%%%%%%%%%%%%%
\tabletypesize{\small}
\setlength{\tabcolsep}{0.05in}
\tablecaption{Point source selection \label{tab4}}
\tablewidth{0pt}
\tablehead{\colhead{ID} & \colhead{Upper boundary} & \colhead{Lower boundary} \\ 
& ($\alpha$, $\beta$, $\gamma$) & ($\alpha$, $\beta$, $\gamma$) }
\startdata
Phot\,A   & 0.03,  0.00023, 22.08 &  --0.03,  --0.00032, 22.15  \\
Phot\,B   & 0.03,  0.00024, 22.15 &  --0.03,  --0.00028, 22.04  \\
Phot\,C   & 0.10,  0.00039, 21.36 &  --0.10,  --0.00072, 22.11  \\
Phot\,D   & 0.10,  0.00036, 21.11 &  --0.10,  --0.00107, 22.23  \\
\enddata
\end{deluxetable} %%%%%%%%%%%%%%%%%%%%%%%%%%%%%

%%%%%%%%%%%%%%%%%%%%%%%
% Figure 6
%%%%%%%%%%%%%%%%%%%%%%%
\begin{figure*}
\centering
\includegraphics[scale=0.6]{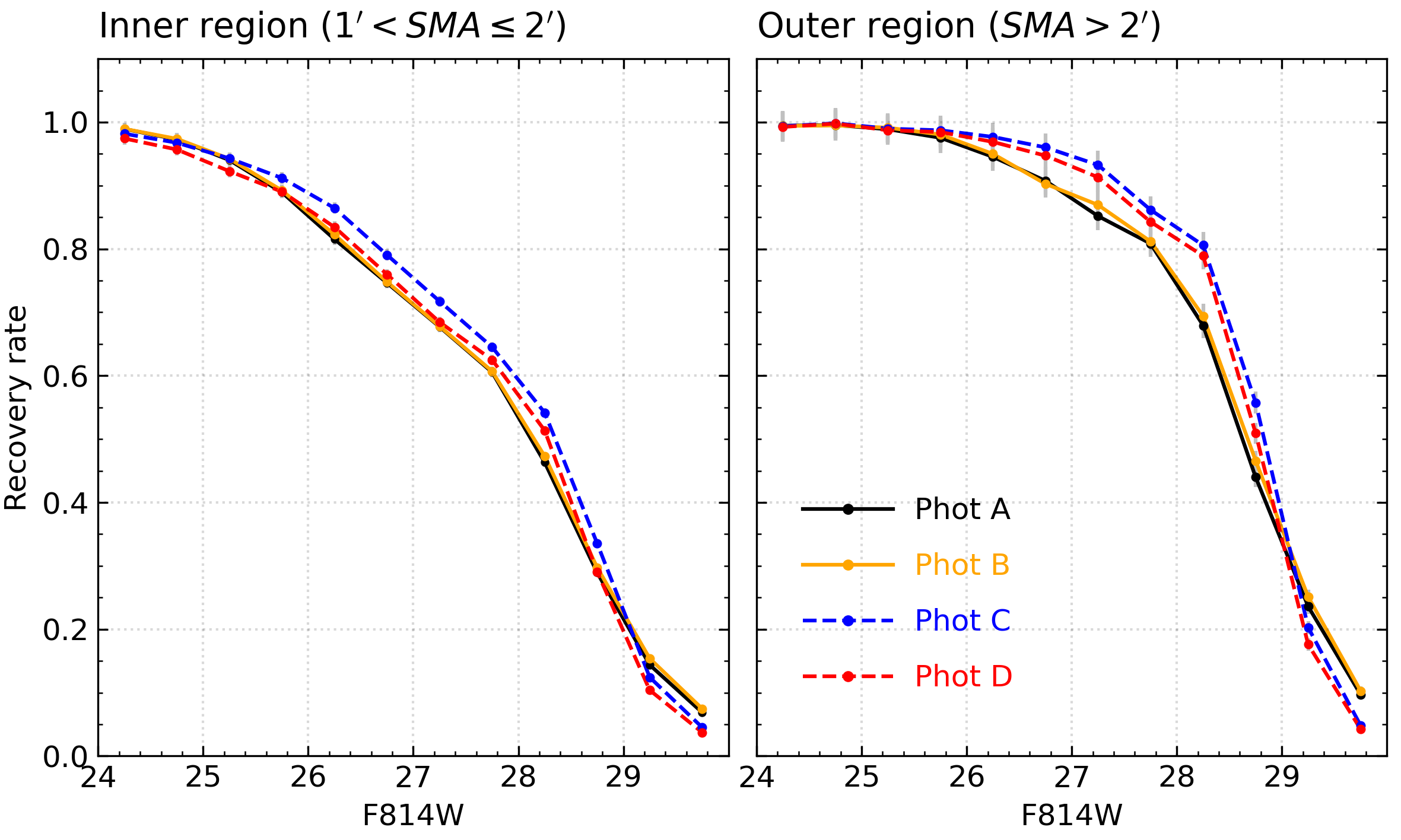}
\caption{
Comparison of photometric completeness (recovery rates) determined from artificial star experiments of individual reductions: Phot\,A -- D. 
Solid and dashed lines indicate the recovery rates taken from the individual-frame photometry (with DOLPHOT) and stacked-frame (with DAOPHOT) photometry, respectively.
Details of the reductions are summarized in Table~\ref{tab1}.
The left and right panels show the recovery rates for the inner ($1\arcmin < SMA \leq 2\arcmin$) and outer ($SMA > 2\arcmin$) regions, respectively.
}
\label{fig6}
\end{figure*}

It is worth mentioning that there is no single setting of point source selection criteria that is optimal for all local environments and all scientific goals.
The selection criteria adopted in this study are designed to sample stars in both the disk and halo simultaneously, and to apply a homogeneous selection (i.e., all sharpness based) to independent reductions as much as possible.
Our approach is rather simple and straightforward compared with the selection schemes in the literature that are often based on several diagnostic parameters (e.g., error, chi-square, roundness, and crowding).
We found, however, that our selection is good enough for our purposes of comparing  photometry  between individual reductions, as discussed in the next section.

\section{Comparison of photometry}
In this section, we used real and artificial star data to assess photometric performance of individual reductions.
We start by analyzing the artificial star data in terms of recovery rates (Section 3.1), statistical errors (Section 3.2), and systematic errors (Section 3.3), which are indicators of completeness, precision and accuracy, respectively.
We then show a star-by-star comparison of real stars and their CMDs (Section 3.4).

%%%%%%%%%%%%%%%%%%%%%%%
% Figure 7
%%%%%%%%%%%%%%%%%%%%%%%
\begin{figure*}
\centering
\includegraphics[scale=0.8]{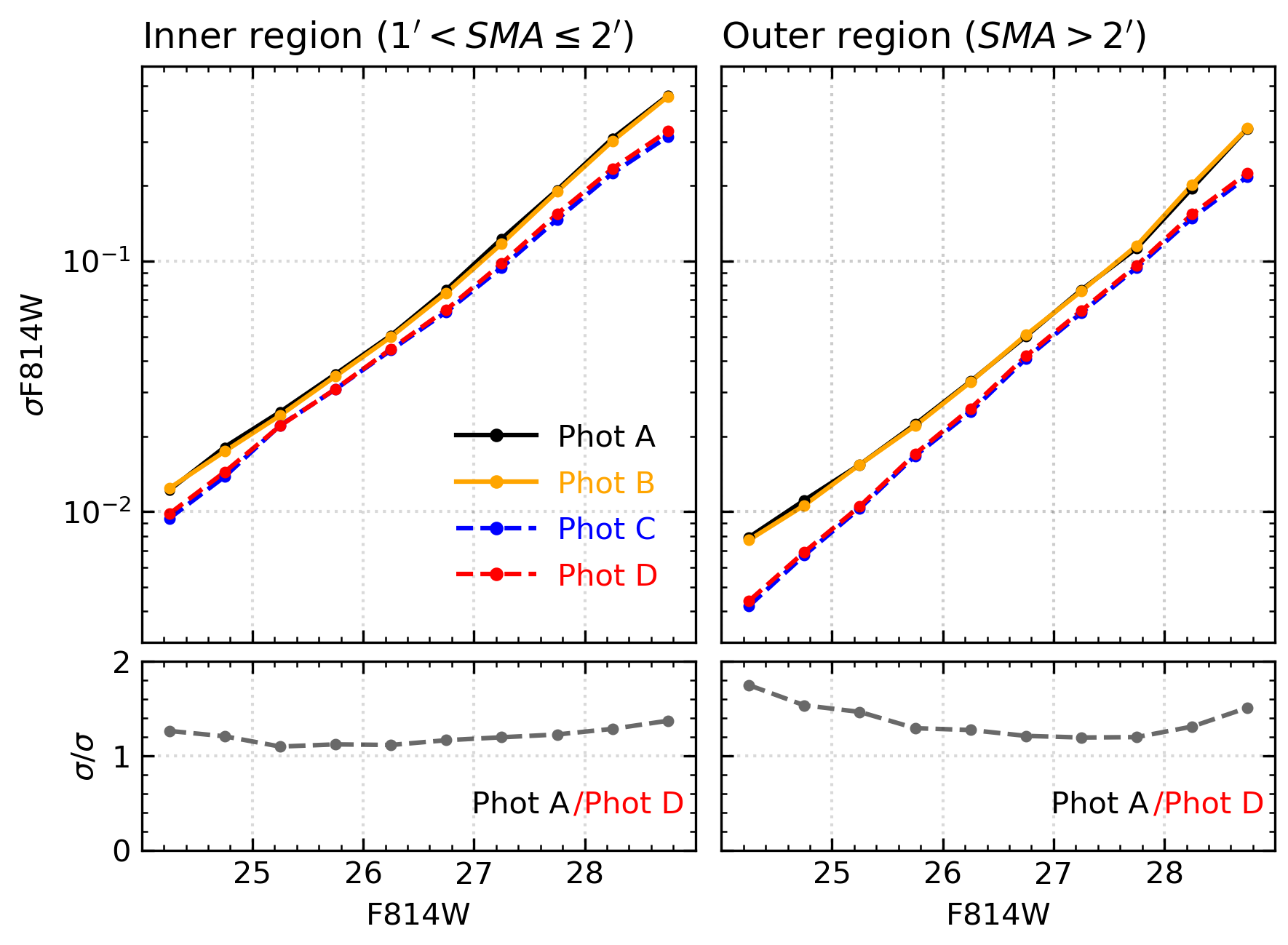}
\caption{Comparison of photometric precision (statistical errors) determined from artificial star experiments.
(Top left) Statistical errors of individual reductions measured from the inner region ($1\arcmin < SMA \leq 2\arcmin$).
(Bottom left) Ratio of errors between Phot\,A (DOLPHOT run on individual frames) and Phot\,D (DAOPHOT run on drizzled frames).
(right) Same as the left panels, but for the outer region ($SMA > 2\arcmin$).
}
\label{fig7}
\end{figure*}

\subsection{Photometric completeness}
In Figure \ref{fig6}, we present completeness of our photometry for the inner (left) and outer (right) regions of the NGC\,3370 field.
The recovery rates taken from individual reductions (Phot\,A -- D) are indicated as a function of input $F814W$ magnitudes. 
Error bars represent the $1\sigma$ statistical uncertainty in each magnitude bin. 
They are mostly smaller than the symbol size.

It is obvious (and expected) that the inner region ($1\arcmin < SMA \leq 2\arcmin$) has lower recovery rates than the outer region ($SMA > 2\arcmin$) regardless of the reduction method.
This is most likely due to the high stellar crowding at the disk region, where the surface brightness is several magnitudes brighter than the outer halo field \citep[e.g.,][]{can09}.
The 80\% recovery rates are measured at $F814W \approx 26.5$ and $\approx28$~mag in the inner and outer regions, respectively.

Individual reductions show notable features in their respective star recovery rates.
First, Phot\,A and B show almost identical completeness curves in both inner and outer regions (solid lines). 
This is explained by the similarity in the two photometry methods:
both Phot\,A and B are based on the DOLPHOT processing with individual frame images.
The only difference is that Phot\,A has aperture correction determined automatically from the code itself ($\tt ApCor=1$), and Phot\,B is based on the manual aperture correction.
The aperture correction typically amounts to a few hundreds of magnitudes, so Phot\,A and B should have similar photometric performance, in recovery rates, statistical and systematic errors.
Phot\,C and D are in the same vein. 
They were reduced using the same code (DAOPHOT) and the same images (drizzled stacked frames). 
A small difference is the PSF models: empirical PSFs for Phot\,C and synthetic TinyTim PSFs for Phot\,D.
The good agreement between their completeness curves (dashed lines) indicates that the choice of PSF models does not meaningfully change the recovery rates.

Second, there is a systematic difference between the two groups of photometry: 
DOLPHOT runs on individual frames (Phot\,A and B) and DAOPHOT runs on stacked frames (Phot\,C and D).
Photometry at the bright side of $F814W \lesssim 25$~mag is almost complete in all the cases.
The systematic difference becomes apparent thereafter.
In the magnitude range of $25.5$ $\lesssim F814W \lesssim 28.5$~mag, Phot\,C and D are more complete, showing $\sim$5\% (up to $\sim$10\%) higher recovery rates than Phot\,A and B. 
This trend is seen in both inner and outer regions.
The completeness curves are crossed over at the faint side ($F814W \gtrsim 29$~mag) showing higher recovery rates for Phot\,A and B than Phot\,C and D.
We note, however, that photometry near the detection limit is easily contaminated by false-positive detections (e.g., sky fluctuations) so that the recovery rates in this magnitude range are less reliable.

In summary, we found a reasonable agreement in photometric completeness between independent reductions. 
While we detected a small systematic difference in some cases, the difference is not observed across the entire magnitude range and doesn't significantly change the general shape of the completeness curves.

%%%%%%%%%%%%%%%%%%%%%%%
% Figure 8
%%%%%%%%%%%%%%%%%%%%%%%
\begin{figure*}
\centering
\includegraphics[scale=0.8]{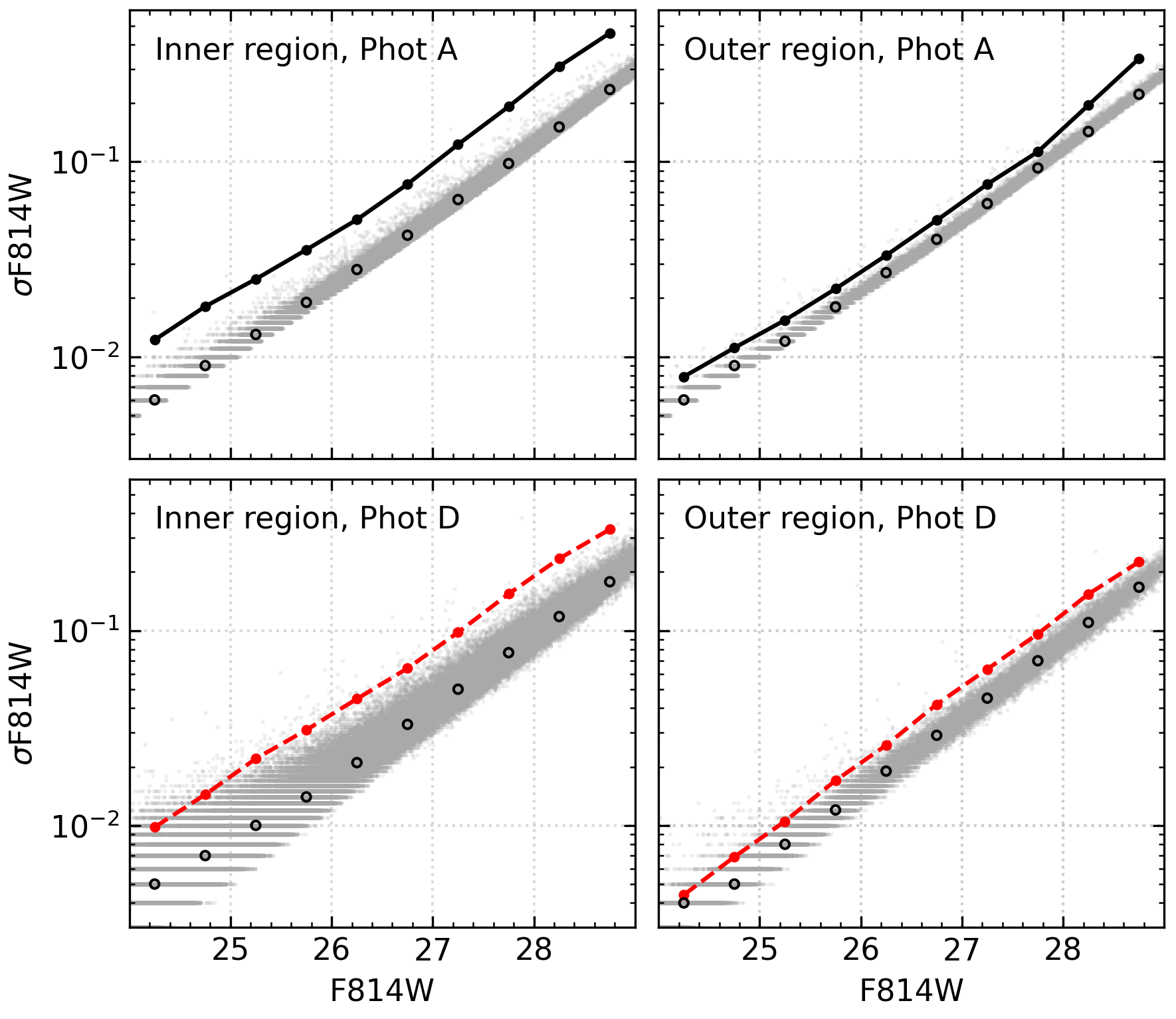}
\caption{Comparison of photometric errors determined from artificial star experiments (circles with lines) and estimated directly from the photometry codes (gray dots and open circles).
Gray dots indicate the errors of individual stars assigned by either DOLPHOT or DAOPHOT. 
Their median errors in each magnitude bin are marked by open circles.
The left and right panels show the results from the inner ($1\arcmin < SMA \leq 2\arcmin$) and outer regions ($SMA > 2\arcmin$), respectively.
We choose two representative reductions: Phot\,A (top panels) and Phot\,D (bottom panels).
}
\label{fig8}
\end{figure*}

\subsection{Photometric precision}
The statistical (random) error of individual reductions were derived from the artificial star data.
We used the standard deviation of the magnitude difference between the injected and recovered stars (e.g., error bars in Figure \ref{fig4}) as a proxy for the statistical errors.
This approach is straightforward and provides robust estimates of photometric errors taking into account both the Poisson photon statistics and stellar crowding \citep[e.g.,][]{ste88, gal96, dol02, mak06}.
 
Figure \ref{fig7} displays the statistical errors for the inner (left) and outer (right) regions of NGC\,3370.
It is not surprising that the inner crowded region shows larger errors than the outer region. 
The difference is about 60\%, which means that the errors of the inner region are on average 1.6 times larger than those of the outer region in the same magnitude bin.

We also found that there is a systematic difference between individual reductions 
such that Phot\,A and B have larger statistical errors than Phot\,C and D. 
To better show the difference, we plot the error ratios between Phot\,A and Phot\,D in the bottom panels.
It is clearly seen that Phot\,A has about 20\% (for the inner region) -- 35\% (for the outer region) larger errors than Phot\,D.
Similar error ratios can be achieved with Phot\,B and C, as their errors are almost the same as those of Phot\,A and D, respectively.
This result indicates that DAOPHOT runs on stacked frames could deliver higher photometric precision (i.e., smaller statistical errors) than DOLPHOT runs on individual frames.

We next explore the statistical errors computed internally from the photometry codes.
Figure \ref{fig8} shows a comparison of errors derived from two independent approaches:
artificial star tests (filled circles with lines) and
internal routines in the photometry codes (gray dots and open circles).
We chose two representative reductions, Phot\,A (top) and Phot\,B (bottom), to see the errors in the inner (left) and outer (right) region of NGC\,3370. 
We found that the internally computed errors are smaller than the errors determined from the artificial star experiments in all cases. 
The average ratio between the two error estimates in the outer region is 1.25 for Phot\,A and 1.38 for Phot\,D.
The inner crowded region shows a more significant difference with larger error ratios: 1.94 for Phot\,A and 1.99 for Phot\,D.
This result indicates that artificial star experiments are critical to properly estimate photometric errors; using the errors computed directly from photometry codes underestimates the true errors, especially in  crowded fields.

%%%%%%%%%%%%%%%%%%%%%%%
% Figure 9
%%%%%%%%%%%%%%%%%%%%%%%
\begin{figure*}
\centering
\includegraphics[scale=0.8]{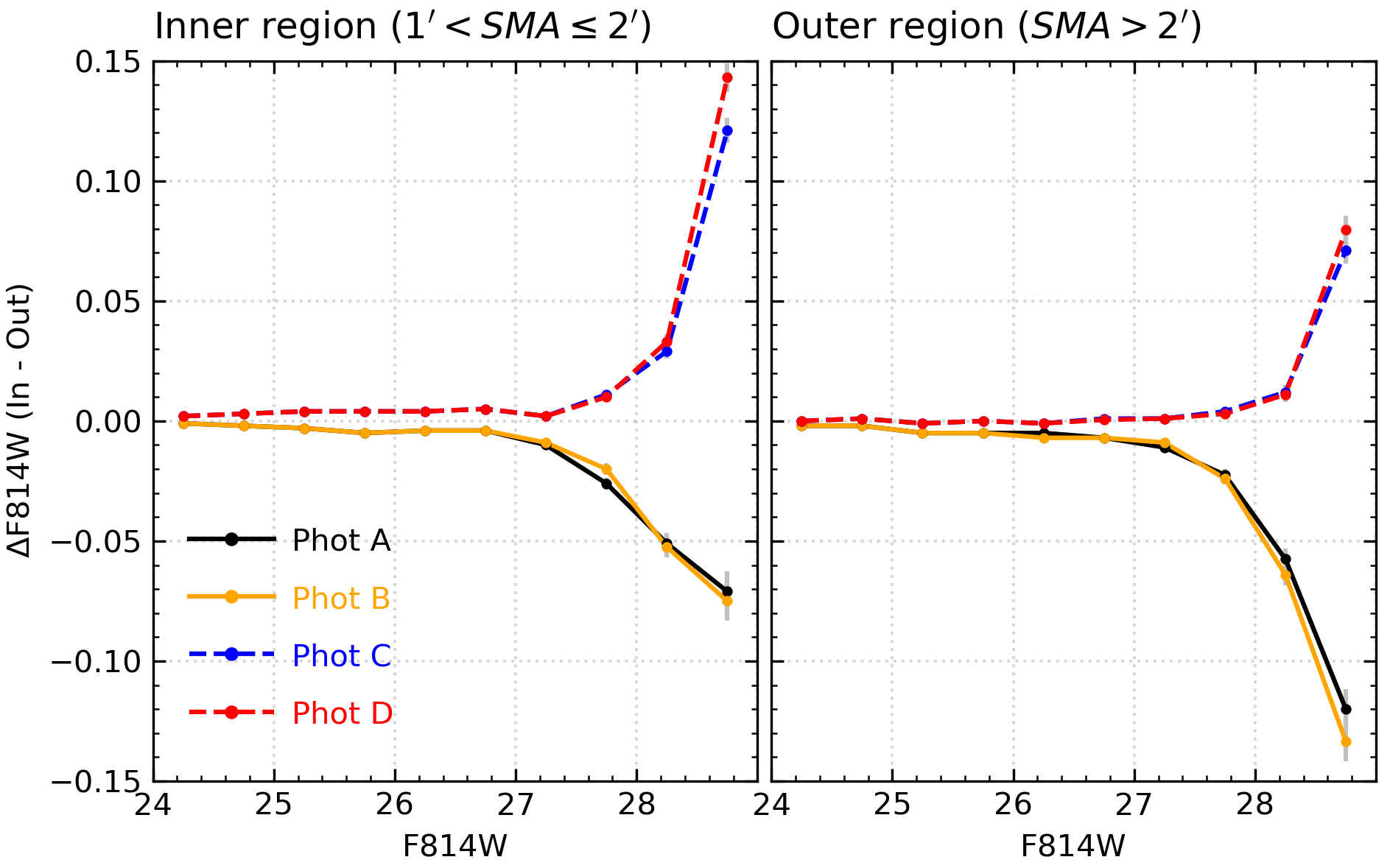}
\caption{Comparison of photometric accuracy (systematic errors) determined from artificial star experiments.
Individual reductions are indicated according to the legend.
The left and right panels show the results from the inner ($1\arcmin < SMA \leq 2\arcmin$) and outer regions ($SMA > 2\arcmin$), respectively.
}
\label{fig9}
\end{figure*}

\subsection{Photometric accuracy}

We estimate the photometric accuracy (systematic error) by examining the magnitude offset between the injected and recovered artificial stars (i.e., red circles in Figure \ref{fig4}).
The results from the inner (left) and outer (right) regions of NGC\,3370 are shown in Figure \ref{fig9}.
A few distinguishable features are evident.
First, the systematic errors for Phot\,A are almost identical to those of Phot\,B in both inner and outer regions.
Similarly, a good match is seen between the errors of Phot\,C and~D.
This, again,
is due to the similarity in data processing, as discussed in previous sections.

Second, there is a stark difference between the two groups of photometry:
DOLPHOT-based reductions with individual frames (Phot\,A and B), and
DAOPHOT-based reductions with stacked frames (Phot\,C and D).
At the faint magnitude end ($F814W \gtrsim 27$~mag), the offsets have opposite signs: 
negative offsets for Phot\,A and B (i.e., fainter recovered magnitudes), and positive offsets for Phot\,C and D (i.e., brighter recovered magnitudes).
The brighter recovered magnitudes is naturally expected at the faint side as unresolved stars below the detection limit will contribute to source's magnitudes.
The origin of the opposite trend for Phot\,A and B is not clear, 
but we infer that this is likely due to the sky estimation in individual frames, where sources have much lower signal to noise ratio than the stacked frames.

Third, the systematic errors at the bright end of the magnitude range ($F814W \lesssim 27$~mag) are almost negligible in both the inner and outer regions.
The errors become larger at the faint end, but do not exceed $|\Delta F814W| = 0.15$~mag down to $F814W = 29$~mag.
The 50\% recovery rates are measured at $F814W \sim 28.2$~mag in the inner region (see Figure \ref{fig6}).
At this level, the inner region gives systematic errors of 
$\sim$0.05~mag for Phot\,A and B, and $\sim$0.03~mag for Phot\,C and D.
The same completeness level for the outer region is measured at $F814W \sim 28.6$~mag, and there we found systematic errors of 
$\sim$0.11~mag for Phot\,A and B, and $\sim$0.07~mag for Phot\,C and D.
It is noted that these systematic errors are well within the statistical errors ($0.2\sim0.3$~mag, see Figure \ref{fig7}).

In summary, we detected measurable systematic errors at the faint magnitude range of photometry.
The sign of the systematic effects is not the same in all the cases. 
The photometric accuracy is not much different between individual reductions, though we found slightly smaller systematic errors for Phot\,C and D than Phot\,A and B in the outer region.

%%%%%%%%%%%%%%%%%%%%%%%
% Figure 10
%%%%%%%%%%%%%%%%%%%%%%%
\begin{figure*}
\centering
\includegraphics[scale=1.0]{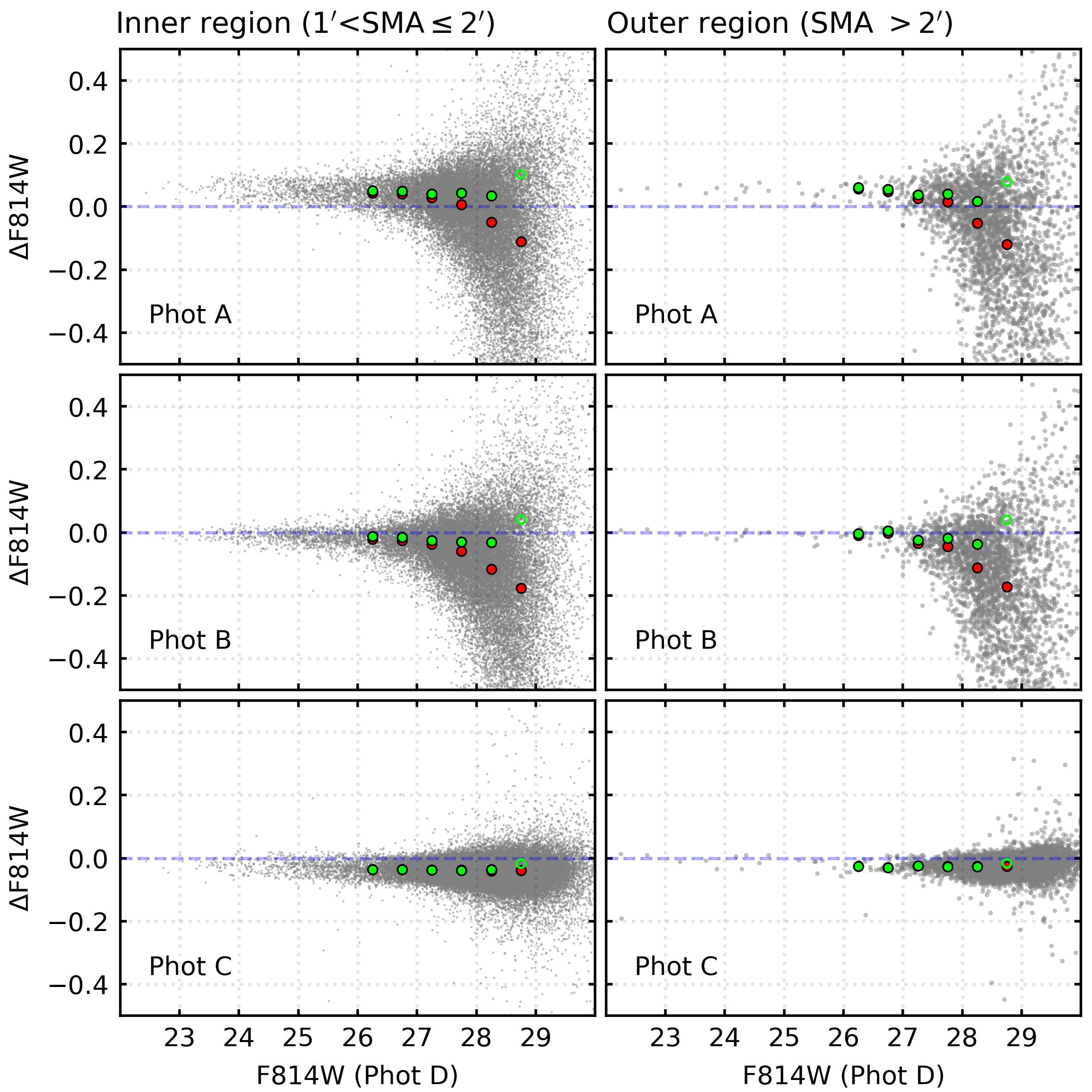}
\caption{Comparison of real-star photometry taken from the inner (left) and outer (right) regions of NGC~3370.
The residual differences between individual reductions are indicated as a function of Phot\,D magnitudes. 
Median offsets in each magnitude bin are marked by red dots. 
Green dots are the median offsets corrected for the systematic bias measured from the artificial star tests. 
The bias correction at the faintest level ($F814W \sim 28.75$~mag) is uncertain, so we marked the corrected offset by open circles.
}
\label{fig10}
\end{figure*}

\begin{deluxetable*}{c|ccc|cccc} %%%%%%%%%%%%%%%%%%%%%%%%%%%%%
\tabletypesize{\small}
\setlength{\tabcolsep}{0.05in}
\tablecaption{Summary of magnitude differences between Phot\,D and other reductions \label{tab5}}
\tablewidth{0pt}
\tablehead{\colhead{Magnitude range} & \multicolumn{3}{c}{Inner region ($1'< SMA \leq 2'$)} & \multicolumn{3}{c}{Outer region ($SMA > 2'$)}\\
($F814W$) & \colhead{Phot\,A} & \colhead{Phot\,B} & \colhead{Phot\,C} & \colhead{Phot\,A} & \colhead{Phot\,B} & \colhead{Phot\,C} }
\startdata
$\leq26$      &   0.050         & --0.015          & --0.029          &   0.051        & --0.008          & --0.010           \\ 
$26.0 - 26.5$ &   0.043 (0.051) & --0.021 (--0.013)& --0.036 (--0.036)&   0.057 (0.061)& --0.010 (--0.004)& --0.025 (--0.025) \\ 
$26.5 - 27.0$ &   0.040 (0.049) & --0.025 (--0.016)& --0.036 (--0.036)&   0.048 (0.056)& --0.003   (0.005)& --0.030 (--0.031) \\ 
$27.0 - 27.5$ &   0.029 (0.041) & --0.037 (--0.026)& --0.037 (--0.037)&   0.026 (0.038)& --0.034 (--0.024)& --0.024 (--0.024) \\ 
$27.5 - 28.0$ &   0.007 (0.043) & --0.060 (--0.030)& --0.038 (--0.039)&   0.015 (0.041)& --0.045 (--0.019)& --0.026 (--0.027) \\ 
$28.0 - 28.5$ & --0.050 (0.034) & --0.117 (--0.032)& --0.040 (--0.036)& --0.052 (0.017)& --0.113 (--0.038)& --0.026 (--0.027) \\ 
$28.5 - 29.0$ & --0.111 (0.102) & --0.177   (0.040)& --0.039 (--0.018)& --0.120 (0.080)& --0.173   (0.040)& --0.025 (--0.017) \\ 
\enddata
%\begin{tablenotes}
%\small
%\item {\bf Note.} Values in parentheses are the offsets corrected for the photometric bias measured from the artificial star experiments.
%\end{tablenotes}
\tablenotetext{a}{Note: values in parentheses are the offsets corrected for the photometric bias measured from the artificial star experiments.}
\end{deluxetable*} %%%%%%%%%%%%%%%%%%%%%%%%%%%%%

\subsection{Comparison of real stars}
With the extensive artificial star data and their photometric properties in hand, we compared real-star photometry.
Figure \ref{fig10} displays the magnitude difference between individual reductions for the inner (left) and outer (right) regions of NGC\,3370.
We used Phot\,D, which is a stacked-frame DAOPHOT photometry with TinyTim PSFs, as a reference reduction to see residual differences ($\Delta F814W$ = Phot\,D -- others).
The red circles indicates the observed median offset in each magnitude bin.
These offsets can be corrected for the systematic bias measured from the artificial star experiments, as marked by green circles.

The brightest stars exhibit a small scatter in $\Delta F814W$ so systematic offsets can be easily identified.
We found that Phot\,A shows larger offsets than Phot\,B and C in both inner and outer regions, as listed in Table \ref{tab5}.
The median offsets for stars brighter than $F814W = 26$~mag in Phot\,A are $\Delta F814W$ = 0.050 and 0.051~mag for the inner and outer regions, respectively.
These values are much larger than the offsets from the other reductions: $\Delta F814W$ = --0.015 and --0.008~mag (inner and outer regions) for Phot\,B and --0.029 and --0.010~mag (inner and outer regions) for Phot\,C.
The larger offsets for Phot\,A are most likely due to the different methodology adopted for the aperture correction: 
Phot\,A uses the automated routines implemented in DOLPHOT ({\tt{ApCor=1}}) and the other reductions are based on the manual determinations with visual inspections of images (see Section 2.3).
According to the DOLPHOT manual, the default aperture correction could have the potential for errors.
We infer that the DOLPHOT processing of the NGC\,3370 field (based on Phot\,A) results in a  significant error in the aperture correction.
This result supports the general conclusion that a careful inspection of the aperture correction is an essential component of obtaining accurate photometry.

Looking at the faintest stars, we found that Phot\,A and B show clear magnitude-dependent offsets.
The observed offsets (red circles) decrease with increasing magnitude and they reach $\Delta F814W \simeq -0.15$~mag in the faintest magnitude bin. 
This is explained by the systematic bias of individual reductions.
Through the artificial star experiments, we confirmed that Phot\,A and B have a negative systematic bias at the faint end (i.e., recovered magnitudes are fainter than the input magnitudes), and the other two reductions (Phot\,C and D) have an opposite trend (i.e., recovered magnitudes are brighter than the input magnitudes).
Therefore, a direct photometry comparison of Phot\,D with Phot\,A and B should show very apparent offsets. 

The observed offsets in the photometry comparison can be corrected using the measure of photometric bias, as marked by green circles in Figure \ref{fig10}.
It is clear that the green circles show reduced magnitude dependent offsets, except for the last point at $F814W = 28.75$~mag.
The larger offset for the last point (open circle) is likely due to the nature of the bias correction; 
the photometric bias is measured as a function of the input magnitudes (Figure\,\ref{fig9}), and the bias correction is applied to the observed magnitudes (Figure\,\ref{fig10}).
Therefore, the bias correction is expected to work well when the degree of the correction is not very significant ($\Delta F814W_{corr} \lesssim 0.1$~mag).
Acknowledging this limitation, we found stable and approximately constant offsets over the entire magnitude range (Table \ref{tab5}).

It is also noted that Phot\,C shows measurable offsets in both the inner and outer regions of NGC~3370.
Phot\,C is 
fainter than Phot\,D.
These two sets of photometry
were analyzed using almost the same processing method, except for PSF images: empirical PSFs for Phot\,C and Tiny Tim PSFs for Phot\,D.
Therefore, the measurable offsets come entirely from the choice of PSF images.

\subsection{Color-Magnitude Diagrams}
\begin{figure*}
\centering
\includegraphics[scale=0.8]{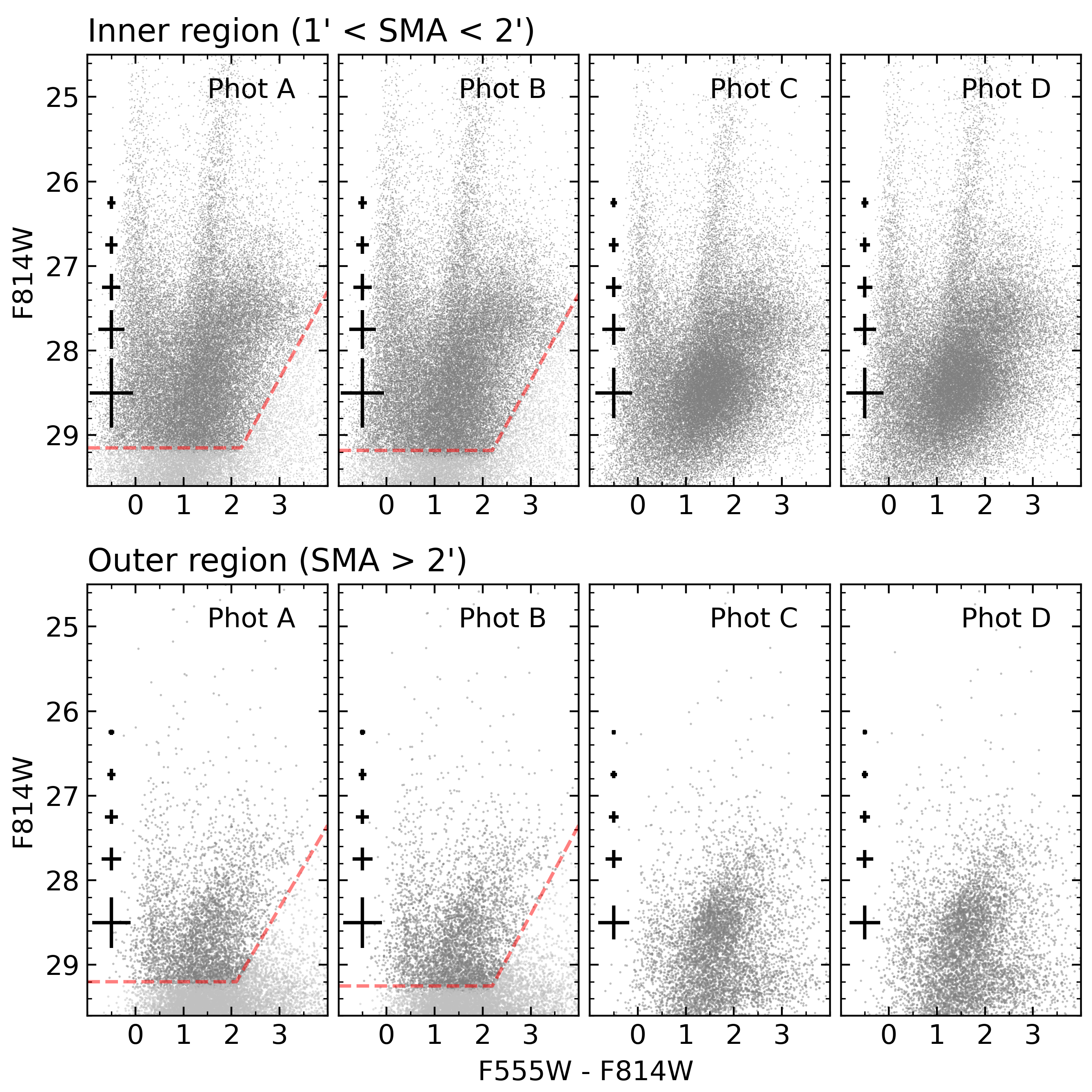}
\caption{
CMDs of resolved point sources in the inner (top) and outer (bottom) regions of NGC~3370. 
The results of individual reductions are shown in each panel.
Photometric errors at $(F555W - F814W) \sim 1.3$ determined from the artificial star experiments  are marked at $(F555W - F814W) = -0.5$.
Dashed lines in the left two panels indicate approximate limiting magnitudes with signal-to-noise ratios of 3 in $F814W$ and 1 in $F555W$.
}
\label{fig11}
\end{figure*}

Figure \ref{fig11} presents
color-magnitude diagrams (CMDs) of resolved point sources in the inner (top) and outer (bottom) regions of NGC~3370 taken from individual reductions.
We used signal-to-noise ratios returned from DOLPHOT to mark an approximate boundary of $S/N = 3$ in $F814W$ and $S/N = 1$ in $F555W$, as shown by a dashed line. 
Sources below this line are less reliable, as they have large (random and systematic) errors with low recovery rates.

CMDs of the inner region show stellar populations with distinct ages: 
young main sequence stars (a vertical feature at $F555W - F814W \approx 0$), 
red helium-burning stars (a slanted feature reaching to $F814W = 24.5$~mag at $F555W - F814W \approx 2$), 
old red giant branch (RGB) stars (a feature below $F814W \approx 28$~mag containing the largest number of stars), and
asymptotic giant branch (AGB) stars (a feature between $F814W \approx 27$ and 28~mag having the reddest color).
These features are seen in all four CMDs from individual reductions. 
We note, however, that the presence of the old RGB population is more evident in Phot\,C and D than Phot\,A and B. 
We infer that this is due to the differences in photometric performance as follows:
1) Phot\,C and D have higher ($\sim$5\%) recovery rates in the RGB magnitude range;
2) Phot\,C and D have smaller ($20\% \sim 30\%$) photometric errors, as marked by error bars in the figure; and 
3) Phot\,C and D have a positive systematic bias at the faint side (i.e., brighter recovered magnitudes). For Phot\,A and B, however, the recovered magnitudes are fainter, so the features in CMDs are more elongated.

Photometry of the outer region is deeper and more precise than the inner region, owing to its lower stellar density (i.e., less crowding).
Indeed, CMDs of the outer region in Figure~\ref{fig11} exhibit a dominant old RGB population in all four panels.
Small photometric errors for Phot\,C and D make them useful to investigate fine structures at the fainter magnitudes in the CMD, such as the tip of the RGB.

%%%%%%%%%%%%%%%%%%%%%%%
% Figure 11
%%%%%%%%%%%%%%%%%%%%%%%

\section{Summary and Discussion}
The distance measurements of nearby galaxies have improved significantly in recent years.
Given the great importance of the local distance scale in the determination of cosmological parameters, a more stringent control of errors is to be desired.
This paper has investigated various systematic issues associated with the data reduction methods, together with presenting a quantitative comparison of photometric performance.

We selected deep observations for NGC\,3370 taken with ACS/WFC on board $HST$ (Figure\,\ref{fig1}). 
The ACS data were reduced using four methods divided into two groups: DOLPHOT runs on individual frames (Phot\,A and B) and DAOPHOT runs on stacked frames (Phot\,C and D) (Table\,\ref{tab1}).
These reductions include different aperture correction methods (automatic vs. manual) and PSF models (synthetic vs. empirical).
Photometric performance (recovery rates, statistical and systematic errors) of individual reduction methods have been  evaluated using artificial star tests.

We have found that photometric incompleteness is comparable in all four reductions (Figure\,\ref{fig6}).
The systematic difference is smaller than 10\% in recovery rates and it does not  change the general shape of the completeness curves.
The overall agreement indicates that the DOLPHOT processing does a good job of detecting faint sources, comparable with the stacked frame photometry with DAOPHOT.

Photometric precision is measured to be 20\% -- 30\% higher for Phot\,C and D (DAOPHOT runs on stacked frames) than Phot\,A and B (DOLPHOT runs on individual frames).
Such a trend is seen in both inner and outer regions of NGC~3370 (Figure\,\ref{fig7}).
We also found that statistical errors assigned to each detection by the photometry codes are significantly smaller than the errors estimated from the artificial star tests (Figure\,\ref{fig8}).
Therefore, artificial star tests are crucial to properly estimate errors of photometry and their impact on the error budget  for subsequent measurements, such as stellar distances and cosmological parameters based upon them (e.g., a local value of $H_0$).

Photometric bias is also detected (Figure\,\ref{fig9}). 
The bias is almost negligible in the bright magnitude range ($F814W \lesssim 27$~mag) but becomes larger at fainter magnitudes, making the recovered magnitudes brighter (for Phot\,C and D) or fainter (for Phot\,A and B).

Due to the photometric bias, a direct comparison of real stars shows large magnitude dependent offsets (Figure\,\ref{fig10}).
These offsets are reduced after applying the bias correction.
Small, but non-negligible systematic offsets ($0.01 \sim 0.03$~mag) are detected in comparisons between Phot\,B, C, and D.
This implies that there could be  reduction-dependent uncertainties remaining after the cited photometric bias correction has been applied.
Similar results can be found in \citet{jan21}. 
They presented photometry of the NGC~4258 halo field with a number of sky fitting options ({\tt Fitsky = 1}, {\tt 2}, and {\tt 3}) and PSF models (TinyTim and Anderson's PSFs) available in DOLPHOT.
A direct photometry comparison showed offsets of $\sim$0.02~mag at the faint side and they too were not entirely eliminated with their bias correction (see their Figures 8 and 9 and A1).
Studies of NGC~3370 and NGC~4258 showed that the reduction dependent uncertainties can be reduced to a 0.02~mag, corresponding to 1\% of the luminosity distance.
We expect that a systematic survey of nearby galaxies will deliver
a more robust estimate of the reduction-dependent uncertainties in the stellar distance scale and the measurement of the Hubble constant.

The automatic aperture correction option ({\tt ApCor=1}) in DOLPHOT has also been tested.
This option is widely used in  DOLPHOT processing, but appears to be subject to larger errors, as documented in the manual.
We have found that there is a large discrepancy ($\sim$0.05~mag) between the automatic and manual aperture corrections (Figure\,\ref{fig10}), indicating that the automatic correction routines applied to the NGC~3370 dataset introduces a sizable systematic error. 
Therefore, the automatic correction option should be used with care and a manual examination is necessary to achieve high accuracy photometry in the DOLPHOT processing.

CMDs of independent reductions are comparable in showing distinct stellar populations in NGC~3370 (Figure\,\ref{fig11}).
One noticeable difference is that Phot\,C and~D CMDs provide  a noticeably  clearer delineation of the old RGB population in both the inner and outer regions.
This implies that the stacked frame photometry is helpful for  studying non-variable stars in external galaxies.

\acknowledgments
I am grateful to Andrew Dolphin for patiently answering my questions over the past few years.
I appreciate Wendy Freedman, Barry Madore, and Myung Gyoon Lee for their careful reviews of the paper.
I thank Kayla Owens for improving the original manuscript.
I thank the SH0ES team for taking deep images of nearby galaxies used in this and my previous studies.
I thank Lucas Macri and Wenlong Yuan for helpful discussions on data reduction.
I appreciate Sungsoon Lim for teaching me how to use DAOPHOT and DOLPHOT in my early career years.
My thanks to Michele Cantiello for useful discussions regarding the stellar structure of NGC~3370.
I would like to thank the members of the CCHP, and GHOSTS teams who
have graciously shared their knowledge and insights into resolved stellar populations. 
My spatial thank to Alisa Brewer, Denija Crnojevic, and Dmitry Makarov.
This research has made use of the NASA/IPAC Extragalactic Database (NED),
which is operated by the Jet Propulsion Laboratory, California Institute of Technology,
under contract with the National Aeronautics and Space Administration.

\end{document}